\documentclass[prl,aps,twocolumn,showpacs]{revtex4}
\usepackage{bm}
\usepackage{epsfig}
\usepackage{amsmath}
\usepackage{color}

\renewcommand{\v}[1]{{\bf #1}}
\newcommand{\bpm}{\begin{pmatrix}}
\newcommand{\epm}{\end{pmatrix}}
\newcommand{\ba}{\begin{eqnarray}}
\newcommand{\ea}{\end{eqnarray}}
\newcommand{\nn}{\nonumber \\}

\begin{document}

\title{Microscopic Theory of Rashba Interaction in Magnetic Metal}

\author{Jin-Hong Park}
\affiliation{Department of Physics and BK21 Physics Research
Division, Sungkyunkwan University, Suwon 440-746, Korea}
\author{Choong H. Kim} \affiliation{Department of
Physics and Astronomy, Seoul National University, Seoul 151-742,
Korea}
\author{Hyun-Woo Lee}
\email[Electronic address:$~~$]{hwl@postech.ac.kr} \affiliation{PCTP
and Department of Physics, Pohang University of Science and
Technology, Pohang, 790-784, Korea}
\author{Jung Hoon Han}
\email[Electronic address:$~~$]{hanjh@skku.edu}
\affiliation{Department of Physics and BK21 Physics Research
Division, Sungkyunkwan University, Suwon 440-746, Korea}
\affiliation{Asia Pacific Center for Theoretical Physics, POSTECH,
Pohang, Gyeongbuk 790-784, Korea}

\date{\today}

\begin{abstract} Theory of
Rashba spin-orbit coupling in magnetic metals is worked out from
microscopic Hamiltonian describing $d$-orbitals. When structural
inversion symmetry is broken,  electron hopping between $d$-orbitals
generates chiral ordering of orbital angular momentum, which
combines with atomic spin-orbit coupling to result in the Rashba
interaction. Rashba parameter characterizing the interaction is
band-specific, even reversing its sign from band to band. Large
enhancement of the Rashba parameter found in recent experiments is
attributed to the orbital mixing of $3d$ magnetic atoms with
non-magnetic heavy elements as we demonstrate by first-principles
and tight-binding calculations.
\end{abstract}

\maketitle

Control of the local magnetization direction in a metallic
ferromagnet through delivery of torque by the current-carrying
electrons is one of the major endeavors of spintronics community
nowadays~\cite{review}. Prompted by the theoretical discovery of a
new type of spin transfer torque (STT) arising in Rashba-coupled
bands~\cite{tatara-Rashba,mz,abiague,Rashba} and subsequent
experimental indication thereof~\cite{miron}, several
proposals~\cite{nonadiabatic-Rashba-STT,Kim12PRB} were made
recently to uncover a complete set of STT's permissible in
ferromagnets with the Rashba-coupled bands, or Rashba ferromagnets
for short.
Experiments to measure Rashba-induced STT were performed for a very
thin atomic layer of magnetic atoms with several layers of heavy
atoms such as Pt grown on top~\cite{miron}. It was predicted that
the Rashba interaction can make the electric control of the
magnetization dynamics drastically more efficient~\cite{Kim12PRB}
and give rise to giant spin motive force~\cite{Kim12PRL}, greatly
raising the technological prospect of spintronic devices based on
unprecedented strong mutual coupling of magnetization dynamics and
electric current.

Compared to other spintronic phenomena such as giant
magnetoresistance and spin Hall effect~\cite{SHE,ralph-2},
theories of the Rashba interaction in ferromagnets are largely
phenomenological~\cite{tatara-Rashba,mz,abiague,nonadiabatic-Rashba-STT,Kim12PRB,Kim12PRL},
based on free electron-like Hamiltonian such as
\ba {\cal H} = {\cal H}_0 + {\cal H}_\mathrm{ex} + {\cal
H}_\mathrm{R}, \label{eq:typical-Rashba}\ea
where ${\cal H}_0= \v p^2 /2m$ gives the dynamics of conduction
electrons in the parabolic approximation, ${\cal H}_\mathrm{ex} = -J
\bm \sigma \cdot \v n$ is the ferromagnetic exchange coupling
between the conduction electron spin $\bm \sigma$ and the localized
moment $\v n$, and ${\cal H}_\mathrm{R} = \alpha_\mathrm{R} (\bm
\sigma \times \v p ) \cdot \hat{z}$ is the Rashba spin-orbit
coupling interaction, with the surface normal to a Rashba
ferromagnet taken as the $\hat{z}$-axis. In these theories the
Rashba parameter $\alpha_\mathrm{R}$ is a purely phenomenological
parameter, and it remains unclear what controls the strength of
$\alpha_\mathrm{R}$. A naive estimation of $\alpha_\mathrm{R}$ based
on  the relativistic effective magnetic field in free electron
picture and consequent Zeeman energy results in  $\sim 10^{-4}$
eV\AA, \ which is several orders of magnitude smaller than
experimental values~\cite{miron,Pi10APL,laShell}. In light of the
technological prospect of Rashba ferromagnets, not to mention the
theoretical importance of the problem itself, we try to establish
the Rashba interaction in ferromagnets on a microscopic level. In
recent papers some of the present authors showed that the Rashba
interaction in non-magnetic bands is essentially a multi-orbital
phenomenon~\cite{kim,orbital-Rashba}. A similar idea was proposed
earlier in Ref.~\cite{hedegard}. As the magnetically polarized bands
also typically exhibit multi-orbital character with most of the
$d$-orbitals involved in the band structure, the multi-orbital
scheme may be brought to bear on the magnetic system as well.

Construction of a microscopic Hamiltonian is done for $t_{2g}$
$d$-orbitals ($d_{xy},d_{xy},d_{zx}$) where magnetism typically
occurs. We assume a square lattice and introduce the tight-binding
Hamiltonian $H_{t_{2g}}$ for electron hopping to nearest neighbor
sites with Slater-Koster parameters $t_1$ and $t_2$ for $\sigma$-
and $\pi$-hopping of the $d$-orbitals.
The inversion symmetry breaking (ISB) about the $xy$-plane
generates additional hopping terms~\cite{ON},
\ba && H_\mathrm{ISB} = 
\gamma\sum_{i,\sigma}[c^{\dag}_{i, xy, \sigma} c_{i + \hat{y},
zx,\sigma} + c^{\dag}_{i, xy, \sigma} c_{i + \hat{x}, yz, \sigma} +
h.c.]\nn
&& - \gamma \sum_{i,\sigma}[c^{\dag}_{i, xy, \sigma} c_{i - \hat{y},
zx, \sigma} + c^{\dag}_{i, xy, \sigma} c_{i - \hat{x}, yz, \sigma} +
h.c.], \label{eq:ON}\ea
dictated by the ISB parameter $\gamma$. Here $i$ denotes the atomic
site and $c_{i,xy/yz/zx,\sigma}$ is the electron annihilation
operator of the $d_{xy}$/$d_{yz}$/$d_{zx}$ orbital with the spin
$z$-component $\sigma$. In addition, magnetic exchange
$H_\mathrm{ex}$ and atomic spin-orbit interaction (SOI) Hamiltonian
$H_\mathrm{SOI}$ are introduced as
\ba H_\mathrm{ex}
=-J\sum_{i,\sigma,\sigma'}{\v C}^\dagger_{i,\sigma}{\v n} \cdot
(\bm \sigma)_{\sigma,\sigma'}{\v C}_{i,\sigma'} , \nn
H_\mathrm{SOI}=\lambda_\mathrm{so}\sum_{i,\sigma,\sigma'}{\v
C}^\dagger_{i,\sigma}{\v L}\cdot (\bm \sigma)_{\sigma,\sigma'} {\v
C}_{i,\sigma'}, \ea
where ${\v C}^\dagger_{i,\sigma}=
(c^\dagger_{i,xy,\sigma},c^\dagger_{i,yz,\sigma},c^\dagger_{i,zx,\sigma})$,
and ${\v L}$ is the $3\times 3$ matrix consisting of the
expectation values of the atomic orbital angular momentum (OAM)
within the $t_{2g}$ orbital space. This completes the microscopic
model we shall study now: $H_\mathrm{TB}= H_{t_{2g}} +
H_\mathrm{ISB} + H_\mathrm{ex} + H_\mathrm{SOI}$~\cite{comment}.

\begin{figure}[ht]
\includegraphics[width=85mm]{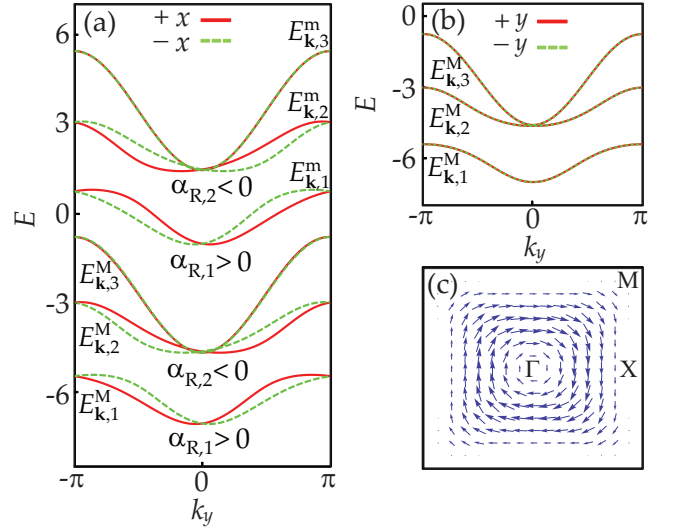}
\caption{(color online) (a) Energy bands for the tight-binding
Hamiltonian $H_\mathrm{TB}$ with $(t_1, t_2, \gamma,
\lambda_\mathrm{so}, J) = ( 1, 0.2, 0.5, 0.5, 3.0)$. Red solid and
green dashed lines are obtained for the unit magnetization vector
$\v n$ directed along $+\hat{x}$ and $-\hat{x}$ directions,
respectively. Majority (M) and minority (m) spin bands are labeled
with superscripts. Signs of the Rashba parameters for the bands
are indicated. The band $a=3$ shows no Rashba effect as indicated
by the absence of band shift. Energy dispersion along the $k_y$
direction are shown.  (b) No discernible difference in the band
structure occurs along the $k_y$ direction for magnetization $\v
n$ directed along $+\hat{y}$ and $-\hat{y}$ directions. (c) Chiral
OAM pattern for $a=1$ majority band with $\lambda_\mathrm{so}=0$.
Other parameters are the same as in (a).}
\label{fig:t2g-dispersion}
\end{figure}

Numerically determined energy bands of $H_\mathrm{TB}$ are shown
in Fig.~\ref{fig:t2g-dispersion}. The result indicates interesting
dependence of the band dispersion on the relative direction of $\v
n$ with respect to the Bloch momentum $\v k$. When $\v n$ is
either parallel or anti-parallel to $\v k$ we do not find any
changes in the dispersion [Fig.~\ref{fig:t2g-dispersion}(b)]. On
the other hand, when $\v n$ is orthogonal to $\v k$, a clear
displacement of the band is observed
[Fig.~\ref{fig:t2g-dispersion}(a)] in the opposite directions
according to whether $\v n
\parallel +\hat{z}\times \v k$ or $\v n \parallel -\hat{z}\times \v
k$ . The amount of the displacement grows linearly with $\v k$
near the $\Gamma$-point ($\v k=0$). We remark that
Fig.~\ref{fig:t2g-dispersion} is obtained in a situation where $J$
is larger than the other energy scales so that the electron spin
direction is essentially parallel or anti-parallel to $\v n$. Then
the behavior of the energy bands in Fig.~\ref{fig:t2g-dispersion}
can be summarized as the energy shift caused by the Rashba
interaction $(\bm \sigma \times \v k) \cdot {\hat{z}}$ in the
presence of $H_\mathrm{ex}$, demonstrating that the
phenomenological Hamiltonian ${\cal H}$
[Eq.~(\ref{eq:typical-Rashba})] can indeed be derived from the
microscopic one, $H_\mathrm{TB}$. This dependence of the energy
bands on the $\v n$ direction has been utilized in the
experiment~\cite{kaindl} to verify the Rashba interaction at Gd
surfaces.

The $\v n$-dependent band shift noted in
Fig.~\ref{fig:t2g-dispersion} vanishes when $\gamma =0$. To capture
the essential role of the ISB parameter in establishing the Rashba
interaction we focus on $H_\mathrm{ISB}=\sum_{\v k}H_\mathrm{ISB}(\v
k)$, written in momentum space through ${\v C}_{{\v
k},\sigma}=N^{-1/2}\sum_i e^{i{\v k}\cdot {\v r}_i}{\v
C}_{i,\sigma}$ [$N$=number of atomic sites]. Near the $\Gamma$-point
we have $H_\mathrm{ISB}(\v k)=-2\gamma a_L \sum_{\sigma} {\v
C}^\dagger_{{\v k},\sigma} ({\v L} \cdot {\v k}\times {\hat {z}})
{\v C}_{{\v k},\sigma}$, where $a_L$ is the lattice constant. Thus
$\v L$ tends to align along the $\pm {\v k}\times {\hat {z}}$
direction. We then combine this tendency with
$H_\mathrm{SOI}=\sum_{\v k}H_\mathrm{SOI}(\v k)$, where
$H_\mathrm{SOI}(\v k)=\lambda_\mathrm{so}\sum_{\sigma,\sigma'}{\v
C}^\dagger_{{\v k},\sigma}[{\v L}\cdot (\bm
\sigma)_{\sigma,\sigma'}] {\v C}_{{\v k},\sigma'}$. Roughly
speaking, the combined effect of $H_\mathrm{ISB}(\v k)$ and
$H_\mathrm{SOI}(\v k)$ is to replace $\v L$ in $H_\mathrm{SOI}(\v
k)$ by ${\v k}\times {\hat {z}}$, producing the Rashba interaction
proportional to $(\v k \times {\hat z})\cdot {\bm \sigma}$.

This heuristic argument can be verified by a perturbation
calculation. In the absence of $H_\mathrm{SOI}$ the bands separate
into three majority ($\bm \sigma \parallel \v n$) and three
minority ($\bm \sigma \parallel -\v n$) eigenstates $|\v
k,a\rangle\otimes |{\bm \sigma}\cdot \v n=\pm\rangle$ ($a=1,2,3$)
of $H_{t_{2g}} + H_\mathrm{ISB} +H_\mathrm{ex}$. Within such band
basis the spin-orbit Hamiltonian $H_\mathrm{SOI}(\v k)$ becomes
\ba H_\mathrm{SOI}(\v k)= \sum_{a=1}^3 \psi^\dag_{\v k, a} {\cal
H}_{\v k, a}\psi_{\v k, a}+ \cdots, \label{eq:U-on-H-soi}\ea
where $\psi_{\v k, a}  = \bpm c_{\v k, a, +\v n} \\
c_{\v k, a, -\v n} \epm$ gives the spinor consisting of the majority
$(\parallel +\v n)$ and the minority $(\parallel -\v n)$ $a$-band,
and $(\cdots)$ gives the inter-band matrix elements which are
ignored. The intra-band $2\times2$ matrix elements is obtained near
the $\Gamma$-point as
\ba && {\cal H}_{\v k, 1} =  -{\cal H}_{\v k, 2} = \alpha_\mathrm{R}
(\v k \times {\hat z})
 \cdot \bm
\sigma, ~~ {\cal H}_{\v k,3} = 0, \label{eq:band-dependent-Rashba}
\ea
where $\alpha_\mathrm{R} \equiv (\lambda_\mathrm{so}a_L) (\gamma /
t_1)$ is the Rashba parameter. Near the $\Gamma$-point the
eigenstates also carry the chiral OAM~\cite{orbital-Rashba}
\ba \langle \v k, 1 |\v L |\v k, 1\rangle = -\langle \v k, 2|\v L |\v
k, 2\rangle \simeq {2\gamma \over t_1} \v k \times \hat{z},
\label{eq:OAM-average}\ea
whereas the eigenstates in the third band does not have
OAM~\cite{orbital-Rashba}. (Results over the full Brillouin zone are
shown in Fig. \ref{fig:t2g-dispersion}(c) for $a=1$ band.)

Putting all the pieces together and returning back to the real
space, we arrive at the effective Hamiltonian truncated within
each majority-minority spin-pair band $a$ as $H_\mathrm{TB}\approx
\sum_a H_a$ with
\ba H_a \!=\! \int d^2 \v r ~\Psi^{\dag}_{\v r,a} \left({\v p^2 \over
2m^*_a} \!-\!J \bm \sigma \cdot \v n \!+\! \alpha_{\mathrm{R},a}
({\bm \sigma}\times \v p)\!\cdot\! \hat{z} \right)\Psi_{\v r,a},
\label{eq:Ha}\ea
where $m^*_a$ is the effective mass of the band $a$. Because of the
large value of $J/\alpha_\mathrm{R}$, the inter-band matrix elements
in Eq.~(\ref{eq:U-on-H-soi}) only makes corrections which are second
order in $\v k$ and are ignored in Eq.~(\ref{eq:Ha}). Now the Rashba
parameter becomes band-dependent, $\alpha_{\mathrm{R},1}
=\alpha_\mathrm{R} = -\alpha_{\mathrm{R},2}$ and even
$\alpha_{\mathrm{R},3} = 0$! Such non-trivial band dependence of the
Rashba parameter cannot be understood from phenomenological
consideration alone. As a result when the magnetization direction
$\v n$ is reversed from $+\hat{x}$ to $-\hat{x}$, the corresponding
shift in the band structure should occur in the opposite directions
along the $k_y$-axis for $a=1$ and $a=2$ and not at all for $a=3$,
as is indeed the case; Fig. \ref{fig:t2g-dispersion}(a). For $\v n
=+\hat{y} \rightarrow -\hat{y}$ the shift in the $k_y$-direction is
minimal; Fig. \ref{fig:t2g-dispersion}(b).

A few remarks are in order. Firstly, upon comparing Eq.
(\ref{eq:band-dependent-Rashba}) to Eq. (\ref{eq:OAM-average}) we
deduce that the sign of the Rashba term in each majority-minority
pair band $a$ is correlated with the chirality of OAM for that
band pair. The OAM non-carrying band $a=3$, in turn, does not have
the Rashba interaction. Both nonzero chiral OAM and Rashba
interaction are consequences of multi-orbital character of the
band and the inversion asymmetry. With the recently developed
circular-dichroism angle-resolved photoemission technique one can
independently probe the orbital chirality of the
bands~\cite{CD-ARPES} and deduce the sign of the Rashba parameter
for the specific band in question. Secondly, the connection
between $H_\mathrm{ISB}$ in Eq.~(\ref{eq:ON}) and the combination
$\v L \cdot (\v k \times {\hat z})$ is natural in view of
symmetry, because the combination violates the inversion symmetry
along $z$-axis and can thus arise generically when ISB occurs. In
other models describing, for instance, $p$-electrons moving in 2D
triangular (relevant for Bi) or square lattice, it can be shown
that the additional hopping Hamiltonian [such as $H_\mathrm{ISB}$
in Eq.~(\ref{eq:ON})] allowed by ISB is also proportional to the
combination near the $\Gamma$-point. To be more strict, a more
general Hamiltonian $\v L \cdot [\v R(\v k) \times {\hat z}]$,
where $\v R(\v k)$ is an arbitrary odd function of $\v k$
consistent with the crystal symmetry, also violates the inversion
symmetry. In fact, it can be verified that $H_\mathrm{ISB}$ in
Eq.~(\ref{eq:ON}) is of this form with $\v R(\v k)=-2\gamma
[\sin(k_x a_L ){\hat x}+\sin(k_y a_L ){\hat y}]$, which reduces to
$-2\gamma a_L \v k$ near the $\Gamma$-point.
Note that $\v R(\v k)\times {\hat z}$ explains the OAM pattern in
the entire Brillouin zone in Fig.~\ref{fig:t2g-dispersion}(c).
Thirdly, although the derivation leading to Eq. (\ref{eq:Ha}) was
done for small SOI, one can equally well solve the large-SOI limit
and obtain the same Hamiltonian as Eq. (\ref{eq:Ha}) in the total
spin $J=1/2$ basis with $\alpha_\mathrm{R} = \gamma
a_L$~\cite{orbital-Rashba}.

Band-to-band variations of $\alpha_{\mathrm{R},a}$ imply that
experimentally determined Rashba parameter~\cite{miron} will be an
average of $\alpha_{\mathrm{R},a}$ in some sense. To clarify the
nature of the averaging, we derive the Rashba-induced adiabatic
STT~\cite{mz} from our approach. To this end we now regard $\v n$ in
Eq.~(\ref{eq:Ha}) as position-dependent and introduce a $\v
r$-dependent SU(2) rotation of the operator $\Psi_{\v r} = U_{\v r}
\psi_{\v r}$ to adjust the spin of the conduction electron to
localized spin, $U^\dag_{\v r} (\bm \sigma \cdot \v n ) U_{\v r} =
\sigma_z$~\cite{tatara-review}. We focus on effects of one
particular majority-minority band pair (subscript $a$ is omitted)
and return to the multiple-band case later. Approximating the
emerging SU(2) gauge potential $-i U^\dag_{\v r} \partial_\mu U_{\v
r}$ by their diagonal components, $-i U^\dag_{\v r}
\partial_\mu U_{\v r} \simeq a_\mu \sigma_z$~\cite{tatara-review},
one arrives at the effective action ${\cal S}$, including the
Berry phase action for spin $\v n$ and the spin Hamiltonian $H_{\v
n}$, as  ($I$=spin size)
\ba && {\cal S} = - 2 I a_L^{-2} \int dt d^2 \v r~ a_0 - \int dt
H_{\v n} \nn
&& +  \int dt d^2 \v r~ \psi^{\dag}_{\v r}[i \partial_t - a_0
\sigma_z] \psi_{\v r} \nn
&& -  \int dt d^2 \v r~ \psi^{\dag}_{\v r}\left( {[\v p + \v a
\sigma_z ]^2\over 2m} - J \sigma_z \right) \psi_{\v r} \nn
&& - \int dt d^2 \v r~ \psi^{\dag}_{\v r}\left(\alpha_\mathrm{R}
 (\v n \times [\v p \sigma_z + \v a ])\cdot \hat{z} \right)
\psi_{\v r}.\ea
Landau-Lifshitz equation of motion for $\v n$ follows
straightforwardly,
\ba && (I a_L^{-2} + \rho^s ) \partial_t \v n + \v n \times {\delta
H_{\v n}\over \delta \v n} \nn
&& + (\v j^\mathrm{s}\cdot \bm \nabla)\v n + 2m \alpha_\mathrm{R} \v
n \times (\v j^s \times\hat{z} ) = 0 . \label{eq:LL}\ea
The spin density takes on the usual form $\rho^s =\psi^\dag \sigma_z
\psi$, while the spin current is modified,
\ba \v j^\mathrm{s} \!=\! {1\over 2m} (\psi^{\dag}\sigma_z [\v p
\psi] \!-\! [\v p \psi^{\dag}]\sigma_z \psi ) \!+\! \left({ \v a
\over m} \!+\! \alpha_\mathrm{R} \hat{z}\times  \v n \right)
\psi^{\dag} \psi, \nn\ea
%
due to the Rashba interaction. The last term on the l.h.s. of Eq.
(\ref{eq:LL}) is the adiabatic Rashba-induced STT discussed in
Refs.~\cite{tatara-Rashba,mz,abiague}

For multiple bands crossing the Fermi level, the terms involving the
spin current in the above equation are generalized to (restoring
band index $a$)
\ba \v j^\mathrm{s}  \rightarrow \sum_a \v j^\mathrm{s}_a \equiv \v
J^s , ~~ \alpha_\mathrm{R} \v j^s  \rightarrow \sum_a
\alpha_{\mathrm{R}, a} \v j^\mathrm{s}_a .\ea
In light of the experimental condition which can only access the
total current, $\v J^s$, it becomes evident that the experimentally
measured $\alpha_\mathrm{R}$ is the effective Rashba parameter
$\alpha_\mathrm{R,eff}$ defined as
\ba \sum_a \alpha_{\mathrm{R}, a} \v j^\mathrm{s}_a  \equiv
\alpha_\mathrm{R,eff} \v J^\mathrm{s} .
\label{eq:effective-alpha}\ea
Due to the band-to-band fluctuation of $\alpha_{\mathrm{R},a}$,
$\alpha_\mathrm{R,eff}$ is a sensitive function of the band
structure and the position of the chemical potential.
Rashba-induced non-adiabatic STT discovered
recently~\cite{nonadiabatic-Rashba-STT,Kim12PRB} cannot be derived
without further including impurity terms in the effective action.

So far our discussion has implicitly assumed magnetic band which
itself is subject to ISB $\gamma$ and atomic spin-orbit coupling
$\lambda_\mathrm{so}$. In conventional $3d$ ferromagnets, an
estimate of the Rashba parameter $\alpha_\mathrm{R} \sim
(\lambda_\mathrm{so} a_L ) (\gamma/t_1)$ yields at most 0.1 eV\AA \
since $\lambda_\mathrm{so}\ll 100$meV typical of $3d$ transition
metals. In contrast, experimental values in the range of 1-2
eV\AA~has been reported in $3d$ magnetic thin films covered with
heavy-element layer~\cite{miron}, which suggests that mixing of
atomic orbitals could result in hybridized bands with both strong
magnetism and large Rashba interaction~\cite{SO-transfer}. In
non-magnetic systems, the enhancement of $\alpha_\mathrm{R}$ through
alloying has been reported~\cite{grioni}.

\begin{figure}[ht]
\centering
\includegraphics[width=85mm]{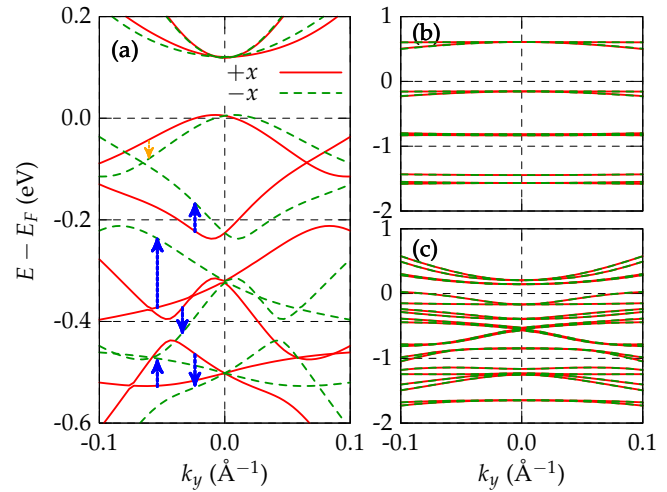}
\caption{(color online) (a) Band structure of 1Co-3Pt layers with
the magnetization forced along $+\hat{x}$ (red full curves) and
$-\hat{x}$ (green dashed curves) directions.  Vertical arrows
connect bands related by Rashba shift. The band associated with
orange arrow is minority band while others with blue arrows are
majority bands. Upward (downward) arrows imply positive (negative)
Rashba parameter. The shift direction is reversed for the minority
band. (b) 1Co layer. (c) 2Pt-1Co-2Pt layer. There is little Rashba
shift observed in (b) and (c) between the two magnetization
directions.}\label{fig:Co-Pt}
\end{figure}

The idea is readily confirmed by the first-principles calculation
with, for example, one Co layer and three Pt layers forming a
heterostructure where the inversion symmetry is naturally
broken~\cite{LDA-comment}. Isolating the eight bands around the
Fermi level, band structures are calculated with the spin
quantization axis constrained along $+\hat{x}$ and $-\hat{x}$
directions, respectively. The energy dispersion measured along the
$k_y$-direction for the two cases are plotted in Fig.
\ref{fig:Co-Pt}(a). The two bands at the top of the figure are
nearly non-magnetic and experience little Rashba shift in response
to the reversal of spin axis. For the remaining six bands which are
magnetic, we deduce significant Rashba parameters
$|\alpha_\mathrm{R}|=1-2$eV\AA~ from the band shifts, in excellent
quantitative agreement with recent experimental values~\cite{miron}.
Composition analysis revealed substantial fraction of both Co and Pt
atoms for all the bands in support of the existence of new
hybridized bands with both Rashba and magnetic characters. Similar
to Fig.~\ref{fig:t2g-dispersion}, the Rashba parameter is heavily
band-specific [Fig.~\ref{fig:Co-Pt}(a)]. In comparison a single
magnetic Co layer yields almost no band shift as seen in Fig.
\ref{fig:Co-Pt}(b). Even imposing perpendicular electric field of 1
eV/\AA~failed to improve the Rashba effect for the Co monolayer. As
another comparison we carried out band structure calculations with
one Co layer sandwiched between two Pt layers on either side to
ensure inversion symmetry. Here we again find almost no Rashba shift
of the band. It can be concluded, then, that twofold conditions are
to be fulfilled for realizing magnetic bands with large Rashba
coupling. One is the hybridization of magnetic and non-magnetic
heavy atom orbitals to ensure the effective mixing of magnetism and
spin-orbit interaction. The second is the inversion symmetry
breaking such as available at surfaces and interfaces to ensure that
Rashba-type interaction becomes symmetry-allowed.

Essential aspects of the mixing can be understood by analysis of
the tight-binding Hamiltonian, $H = H^{\mathrm{NM}} +
H^{\mathrm{M}} + H^{\mathrm{NM-M}}$, consisting of heavy
non-magnetic (NM) bands with Rashba interaction, light magnetic
(M) band, and their coupling (NM-M), respectively. Through the
coupling, each Hamiltonian $H_a$ in the non-magnetic Rashba band
acquires a second-order correction in the hybridization parameter
$t_a$ between non-magnetic $a$-band and the magnetic band,
\ba  H_a &\simeq&  \sum_{\v k} \! \psi^{\dag}_{\v k,
a}\Big(\varepsilon_{\v k, a} \!+\! \alpha_{\mathrm{R},a} ({\bm
\sigma} \times\v k)\cdot \hat{z} \nn
&& ~~~~~ -{|t_a |^2 \over (\varepsilon_{\v k, a} \!-\!
\varepsilon_{\v k, \mathrm{M}})^2 } J \bm \sigma \cdot \v n
\Big)\psi_{\v k, a} .\label{eq:modified-a-band}\ea
Correction to the magnetic band is
\ba H^\mathrm{M} &\simeq& \sum_{\v k} \psi^{\dag}_{\v k, \mathrm{M}}
\Big(\varepsilon_{\v k, \mathrm{M}} \!-\!J \bm \sigma \cdot \v n \nn
&& +\sum_a {| t_a |^2 \alpha_{\mathrm{R},a} \over (\varepsilon_{\v
k, a} \!-\! \varepsilon_{\v k, \mathrm{M}})^2 } ({\bm \sigma}
\times \v k) \cdot \hat{z}\Big) \psi_{\v k, \mathrm{M}}.
\label{eq:modified-M-band}\ea
Apart from some energy dependencies in the denominator, both
Hamiltonians (\ref{eq:modified-a-band}) and
(\ref{eq:modified-M-band}) have remarkable resemblance to the
effective Hamiltonian previously derived, Eq. (\ref{eq:Ha}). Based
on insights from these perturbative calculations we claim the
bands shown in Fig. \ref{fig:Co-Pt}(a) are at once magnetic and
carry a substantial Rashba parameter by virtue of the heavy mixing
of magnetic and non-magnetic orbitals. The magnetization dynamics
of the hybridized bands is also governed by Eqs. (\ref{eq:Ha})
through (\ref{eq:effective-alpha}).

The present investigation showed how magnetic bands with substantial
Rashba interaction can arise in mixed magnetic and heavy-element
structures. In contrast to standard ``relativistic" picture of
Rashba effect, multi-orbital nature and atomic hybridization are
crucial factors in forming the enhanced Rashba magnetic band.
Currently available experimental values of $\alpha_\mathrm{R}$ are
very much scattered~\cite{miron,Pi10APL,ralph-1} even for apparently
similar magnetic layer structures. We speculate that the sensitive
band dependence of the effective Rashba parameter as discussed in
Eq.~(\ref{eq:effective-alpha}) may be partially responsible for
this. Away from the $\Gamma$-point the simple linear-$\v k$
dependence of the Rashba interaction should be replaced by some odd
nonlinear function of $\v k$, which will modify the angular
dependence of Rashba-induced STT as well.

\acknowledgments H. J. H. is supported by NRF grant (No.
2010-0008529, 2011-0015631). H. W. L. is supported by NRF grant
(No. 2010-0014109, 2011-0030789). Informative discussions with
Changyoung Kim and Dongwook Go are acknowledged.


\begin{thebibliography}{24}
\bibitem{comment} This Hamiltonian is also a model of anomalous Hall
effect, as shown in Ref. \onlinecite{ON}.
\bibitem{LDA-comment} For the first-principles density functional theory
(DFT) calculations, we used the Ceperley-Alder parametrization of the
local spin density approximation (LSDA) and projector augmented wave
(PAW) potentials with the Vienna Ab-initio Simulation Package
(VASP)\cite{vasp}. The spin-orbit coupling was included
self-consistently in the non-collinear DFT formalism. The lattice
constant of Co in the hcp structure $a_\mathrm{Co} = 2.507$\AA~ was
assumed for both Co and Pt layers.
\end{thebibliography}
\end{document}